\documentclass[10pt,letterpaper]{article}
\usepackage[top=0.85in,left=2.75in,footskip=0.75in,marginparwidth=2in]{geometry}
\usepackage{threeparttable}
\usepackage{booktabs}
\usepackage{cite}
\usepackage[utf8]{inputenc}
\usepackage{nameref,hyperref}
\hypersetup{
    citecolor=black,
    colorlinks=true,
    linkcolor=blue,
    filecolor=magenta,      
    urlcolor=cyan,
    pdftitle={Overleaf Example},
    pdfpagemode=FullScreen,
    }
\usepackage{microtype}
\usepackage{changepage}
\usepackage[aboveskip=1pt,labelfont=bf,labelsep=period,singlelinecheck=off]{caption}
\usepackage{lastpage,fancyhdr,graphicx}
\usepackage{epstopdf}
\usepackage{color}
\usepackage{graphicx}
\usepackage[pscoord]{eso-pic}
\usepackage[fulladjust]{marginnote}
\usepackage{sidecap}
\usepackage{wrapfig}
\usepackage{nth}
\usepackage{array}
\usepackage{amsmath}
\DisableLigatures[f]{encoding = *, family = * }

\raggedright
\makeatletter
\renewcommand{\@biblabel}[1]{\quad#1.}
\makeatother

\fancyheadoffset[L]{2.25in}
\fancyfootoffset[L]{2.25in}

\definecolor{Gray}{gray}{.25}

\reversemarginpar

\begin{document}
\vspace*{0.35in}

\begin{flushleft}
{\Large
\textbf\newline{Checking the Statistical Assumptions Underlying the Application of the Standard Deviation and RMS Error to Eye-Movement Time Series: \\
\vspace{0.5cm} \hspace{1cm} \large{A Comparison between Human and Artificial Eyes}}
}
\newline
\newline
\\
Lee Friedman \textsuperscript{1}*
Timothy Hanson \textsuperscript{2},
Hal S. Stern\textsuperscript{3},
Oleg V. Komogortsev\textsuperscript{1}
\\
\bigskip
\bf{1} Department of Computer Science, Texas State University,  San Marcos, Texas, USA
\\
\bf{2} Medtronic, Fridley, Minnesota, USA
\\
\bf{3} Department of Statistics, University of California - Irvine, Irvine, CA, USA
\\
\bigskip
* lfriedman10@gmail.com

\end{flushleft}

\section*{Abstract}
Spatial precision is often measured using the standard deviation (SD) of the eye position signal or the RMS of the sample-to-sample differences (StoS) signal during fixation.  As both measures emerge from statistical theory applied to time-series, there are certain statistical assumptions that accompany their use. It is intuitively obvious that the SD is most useful when applied to unimodal distributions.  Both measures assume stationarity, which means that the statistical properties of the signals are stable over time. Both metrics assume the samples of the signals are independent.  The presence of autocorrelation indicates that the samples in the time series are not independent. We tested these assumptions with multiple fixations from two studies, a publicly available dataset that included both human and artificial eyes (“HA Dataset”, N=224 fixations), and data from our laboratory of 4 subjects (“TXstate”, N=37 fixations).  Many position signal distributions were multimodal (HA: median=32\%, TXstate: median=100\%).  
No fixation position signals were stationary.  All position signals were statistically significantly autocorrelated ($p~<~0.01$).  Thus, the statistical assumptions of the SD were not met for any fixation.  All StoS signals were unimodal. Some StoS signals were stationary (HA: 34\%, TXstate: 24\%). Almost all StoS signals were statistically significantly autocorrelated ($p~<~0.01$). For TXstate, 3 of 37 fixations met all assumptions.  Thus, the statistical assumptions of the RMS were generally not met. The general failure of these assumptions calls into question the appropriateness of the SD or the RMS-StoS as metrics of precision for eye-trackers.


\section{Introduction}
\label{sec:intro}
The standard deviation (SD) and the RMS error are metrics which emerge from statistical theory.  As applied to time series, the statistical assumptions underlying their use are that the distributions of samples are unimodal, that time series is stationary, and that the samples in the series are independent. 
In a recent manuscript \cite{Multimodality} we focused on the frequency distributions of horizontal and vertical position signals during fixation, and we noted that most were multimodal. For the present report, we wanted to determine if these time-series were stationary and contain independent samples. A signal is stationary if its statistical properties (median, variance, temporal autocorrelation) do not change over time.  Samples are dependent if autocorrelation exists.  

When we refer to stationarity, we are always referring to “weakly stationary” \cite{Brockwell} or “wide-sense stationarity” \cite{Stationarity}. To be considered stationary, the statistical properties up to the \nth{2} order cannot vary over time. A signal can be median-stationary, which signifies that the median is not changing over time. A signal can also be variance stationary and autocorrelation stationary. Although there are many tests of stationarity \cite{Brockwell}, we chose a relatively new approach that directly determines median and variance stationarity \cite{Stationarity}. We developed our own test of autocorrelation stationarity. Zhivomirov and Nedelchev \cite{Stationarity} explain some of the drawbacks of more traditional tests of stationarity   The proposed method is as follows:  Let's imagine a time-series of 100 samples. The method divides the data in to 2 time-series: one that starts at sample 1 and ends at sample 50, and one that starts at sample 51 and ends at sample 100. It compares the median of the \nth{1} time-series with the median of the \nth{2} time-series using a Mann-Whitney U-test. It compares the variance of the \nth{1} time-series with the variance of the \nth{2} time-series using the Brown-Forsythe Test. It also compares the autocorrelation of the 2 time-series with a method described below.  (The original version of the method called for an initial detrending of the data. We removed that step because we wanted to reject fixations with a trend as non-stationary). The original version of the test only uses a single (50\%-50\%) split of the original time-series, but we make the test stricter by implementing additional splits, as detailed below.   

The temporal autocorrelation of a time-series can be easily tested using an autocorrelation test. This test computes correlations between a time-series and lagged versions of itself. We will show that there is very strong temporal autocorrelation in both our position and StoS fixation time-series. The presence of autocorrelation means that the individual time samples are not statistically independent. Both the SD and the RMS are statistical metrics, and both assume that each sample is independent. Autoregressive time-series models can remove temporal auto-correlation from a univariate time-series: the residuals of such models are statistically independent with no temporal autocorrelation. 

Recently, a dataset became available in which eye movement recordings were collected using both artificial eyes and human eyes during fixation \cite{SmallHead}. Included were data with eye-movement filters off or on during data collection. We thought that this dataset would allow us to address the statistical assumptions of the position signal used for the SD and the StoS signal used for the RMS \cite{HolmqvistBook} in both artificial and human eyes. 

The human subjects in the original dataset \cite{SmallHead} were very good ``fixators’’.  Their fixations were unusually steady with few saccades.  Also, several unusual steps were taken to modify the environment to enhance fixation performance (see below).  To compare these human subjects with more typical human subjects, with more typical environmental conditions, we also analyzed some fixation data from our laboratory.

\section{Materials and Methods}
\label{sec:Methods}
We analyze two datasets in this report.  The one which includes both human and artificial eyes we will refer to as the HA dataset \cite{SmallHead}. The dataset from our laboratory, which includes other human subjects, will be referred to as the TXstate dataset. 

\subsection{The HA Dataset}
\label{sec:HADatasetMethods}
The details regarding the HA dataset were fully explained in the original article \cite{SmallHead}\footnote{The data are available at \url{https://osf.io/search}. Search for ``small head''.}.  In that paper, which included 246 separate recordings, there were many different types of recordings: 5 eye-trackers, 3 different artificial eye types, head fully stabilized and not stabilized, etc.… In the present report we were only interested in studies involving the EyeLink 1000+, with the head stabilized, and we used only the best artificial eye (the SMI artificial eye). Of these 246 studies we employed only 8 studies.  Since artificial eyes cannot be calibrated directly, human calibrations were saved and subsequently used to calibrate artificial eyes. There were two humans (KE and SA). For each human and associated artificial eye, there were recordings with the EyeLink 1000+ filters off or set to “Extra”. The sampling rate was 500 Hz. The characteristics of the 8 studies we analyzed are listed in Table 1.

\begin{table*}[htbp]
\label{tab:tab1}
\begin{adjustwidth}{-1.5in}{0in}
\centering
    \centering
    \begin{threeparttable}
    \caption{Fixation Block Characteristics}
    \begin{tabular}{|c|c|c|c|c|c|c|c|}
        \multicolumn{8}{c}{\large{\textbf{HA Dataset}}} \\ \hline
        Subject & Subject & Filter & N Fixations & N. Excl & Min. Ampl. of & Mdn. Ampl.  & Mdn Fixation\\ 
        Type    &         &        &          & Events \tnote{1} & Excl. Events \tnote{2}& Excl.Events \tnote{3}& Duration (s) \tnote{4}\\ \hline
        Human & KE & Off & 17 & 39 & 0.83 & 1.8 & 3.85 \\ \hline
        Human & KE & Extra & 24 & 36 & 0.86 & 1.89 & 1.88 \\ \hline
        Human & SA & Off & 33 & 108 & 0.36 & 0.61 & 1.48 \\ \hline
        Human & SA & Extra & 38 & 88 & 0.19 & 0.44 & 1.21 \\ \hline
        Artificial & KE & Off & 17 & ~ & ~ & ~ & ~ \\ \hline
        Artificial & KE & Extra & 24 & ~ & ~ & ~ & ~ \\ \hline
        Artificial & SA & Off & 33 & ~ & ~ & ~ & ~ \\ \hline
        Artificial & SA & Extra & 38 & ~ & ~ & ~ & ~ \\ \hline
        \multicolumn{8}{c}{} \\
        \multicolumn{8}{c}{\large{\textbf{TXstate Dataset}}} \\ \hline
        Subject & Subject & Filter & N Fixations & N. Excl & Min. Ampl. of & Mdn. Ampl.  & Mdn Fixation\\ 
        Type    &         &        &          & Events \tnote{1} & Excl. Events \tnote{2}& Excl.Events \tnote{3}& Duration (s)\tnote{4} \\ \hline
        Human & A & OFF & 2 & 2 & 0.7 & 0.74 & 7.25 \\ \hline
        Human & A & ON & 5 & 8 & 0.06 & 0.73 & 2.9 \\ \hline
        Human & B & OFF & 2 & 2 & 0.64 & 1.1 & 7.39 \\ \hline
        Human & B & ON & 6 & 10 & 0.19 & 0.26 & 1.6 \\ \hline
        Human & C & OFF & 2 & 6 & 0.4 & 1.06 & 6.92 \\ \hline
        Human & C & ON & 7 & 6 & 0.3 & 0.39 & 1.22 \\ \hline
        Human & D & OFF & 8 & 14 & 0.18 & 0.31 & 1.47 \\ \hline
        Human & D & ON & 5 & 10 & 0.09 & 0.48 & 1.96 \\ \hline
    \end{tabular}
        \begin{tablenotes}
        \item[1] Number of excluded events
        \item[2] Minimum amplitude of excluded events
        \item[3] Median amplitude of excluded events
        \item[4] Median fixation duration in seconds
    \end{tablenotes}

    \end{threeparttable}
    \end{adjustwidth}
\end{table*}

Each recording contained approximately 100 seconds (50,000 samples) of eye position data.  For the human data it appears that fixations were comprised of repeated periods of approximately 10 seconds, followed by a 5 second break in which no position data were recorded.  Although the data consisted of left and right eye horizontal and vertical position data, in the present report, to keep the number of tables and figures at reasonable levels, only the left horizontal position data were analyzed (both human subjects were left-eye dominant).

Several steps were taken to ensure stable fixations.  All recordings took place in a windowless room with no exterior walls. The room had a constant illumination of 145 lux, measured at the point where the artificial eyes were mounted. Air ventilation was turned off and in-room temperature kept stable at $21^o$ Celsius. The eye trackers and the mounting of artificial eyes were placed on a 300 kg vibration-free table. As soon as the recording started and until it ended, no one was allowed to move in the room. All recordings were collected when no human or machine activity took place in any neighboring room. There were no large roads or manufacturing or construction activity nearby that could introduce vibrations.
Subject KE had 25 years of experience of eye tracking, light blue eyes and used no visual aids. SA had one year of experience of eye tracking, brown eyes and used no visual aids. Both humans had left dominant eyes and upward/forward directed eye lashes. No make-up was used. Both KE and SA exercise regularly, have good muscle control, and can sit very still.

There are 3 types of signal filtering that can be chosen for the EyeLink 1000+: (1) no filtering, (2) “STD” filtering and (3) “Extra” filtering.  The key aspects of the EyeLink filtering algorithm are described in \cite{Stampe}. The filters are heuristic/logical filters (not conventional low-pass filters). The STD filter takes sample N-1, N and N+1 as its input and essentially flattens 1 sample spikes (e.g., physiologically impossible occurrences) in which, for example gaze shifted to the right from N-1 to N and then shifted left from N to N+1). The spikes are flattened by substituting the spike value with the value of its nearest neighbor. The EXTRA filter takes the STD filtered data as its input and is applied to N-2, N-1, N, N+1, N+2, and effectively removes any 2 sample spikes. So, EXTRA incorporates STD.

\subsection{The TXstate Dataset}
\label{sec:TXstateMethods}
The dataset was collected from 4 human subjects (labelled A, B, C and D). There were 3 males and 1 female with ages between 23 and 27. Three were right-eye dominant and one was left-eye dominant.  Only the dominant eye was recorded.

The eye-tracker was an earlier version of the EyeLink 1000+ and is referred to as an EyeLink 1000.  Here is what an expert at SR-Research (the vendors of both EyeLink eye trackers) said about the two systems:
\begin{quote}
“The two systems employ different camera sensors (the Plus has a larger sensor with more pixels) so you may notice that precision metrics are slightly better for the Plus (although this may not be noticeable with real eyes be-cause so many other factors play a role). In general, differences in setup / participants etc… tend to have a greater influence on data quality metrics than hardware differences between the two systems.” (personal communication)
\end{quote}
Data were collected at 1000 samples per second but were down sampled to 500 samples per second (skipping every odd-numbered sample). Stimuli were presented to the subject on a 1680 x 1050-pixel (474 x 297 mm) ViewSonic (ViewSonic Corporation, Brea, California, USA) monitor. Subjects were seated 550 mm in front of the display.

\begin{figure*}[htbp]
\label{fig:fig1} 
\includegraphics[width=12cm]{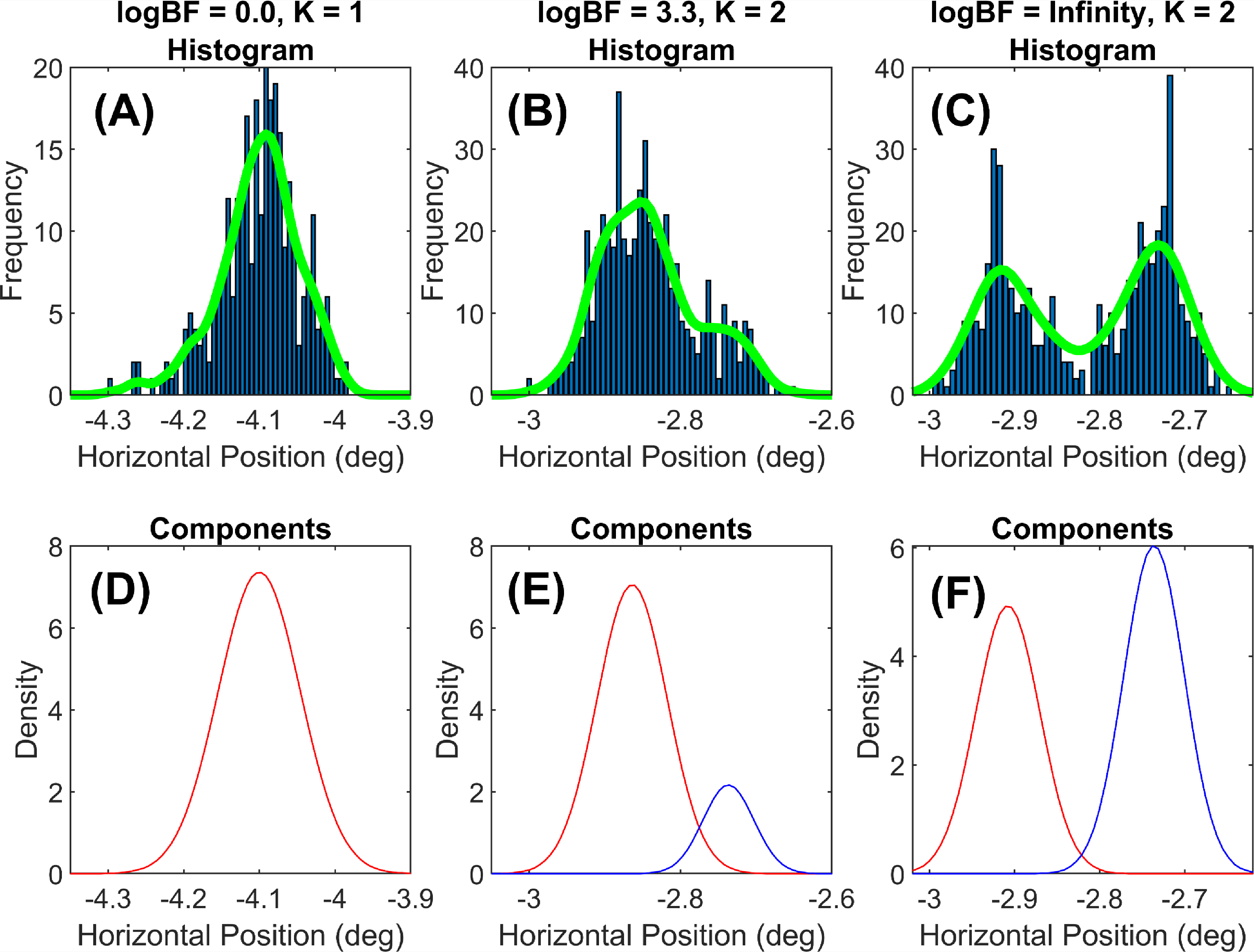}
\caption{Illustration of some multimodality results. All figures based on left eye, horizontal position. (\textbf{A}) Frequency distribution of horizontal position in a fixation with a log(BF) = 0 (unimodal). The density of the histogram is shown in green. (\textbf{B}) Same as (\textbf{A}) but for a fixation with a log(BF) = 3.3 (multimodal). (\textbf{C}) Same as A but for a fixation with a log(BF) that is infinitely large (strongly multimodal). (\textbf{D}) Normal component fitted to data in (\textbf{A}) based on our Bayesian analysis of multimodality. (\textbf{E}) Two normal components fitted to data in (\textbf{B}). (\textbf{F}) Two normal components fitted to data in (\textbf{C}).}
\end{figure*}

\subsection{Data Processing Steps}
\label{sec:DataProcessing}
Data from both datasets underwent identical processing.  Since we were only interested in fixations, all non-fixation events (saccades, microsaccades, full and partial blinks and other artifacts) were removed by manual editing by the first author: he is a very experienced eye-movement researcher who has been involved in eye movement research for 30 years and has published 18 peer-reviewed articles on the topic of eye-movements.  He is the first author on a recent article introducing a comprehensive set of manual eye-movement classification rules \cite{Agreement}\footnote{For 191 examples of events that were removed from these recordings, see: \url{https://digital.library.txstate.edu/handle/10877/15303}. Download ``ExamplesOfEventsManuallyRemoved.zip'' }. Henceforth, these sections will be referred as “removed events”. The signal during these removed events was set to Not-a-Number (NaN)\footnote{All final fixations used in this study are at: \url{https://digital.library.txstate.edu/handle/10877/15303}. Download ''ALL\_FIXATION\_BLOCKS\_HA.zip'' and ``ALL\_FIXATION\_BLOCKS\_Txstate.zip''.}.  

Some characteristics of the removed events are provided for both datasets in Table 1. 

For each fixation, for all 16 recordings (8 from the HA dataset and 8 from the TXstate dataset), the following metrics were calculated for both position signals and StoS signals: (1) a Bayes Factor (BF) indicating unimodality or multimodality, (2) an assessment of the stationarity of each fixation (present or absent), and (3) the lag 1 temporal autocorrelation, which we will refer to as “ACF 1”. The ACF 1 is simply the correlation between each sample and the previous sample. The unimodality assessment, the stationarity assessment and the autocorrelation assessment are explained below.

\subsection{Assessing Unimodality}
\label{sec:Unimodality}
To determine if the distributions of position signals in each fixation were unimodal or multimodal, we employed the Bayesian mixture model approach described in \cite{Xu} (see Figure 1 for an illustration of this process).
 
The basic idea is that an algorithm is employed to try to fit from $1 \dotsb kmax$ (5, in our case) weighted normal distributions to the histogram of the horizontal position signals for each fixation. Each normal component is represented by a mean, a standard deviation (SD), and a weight. 

The sum of these weights is always 1. This is done repetitively, 2000 times (iterations), and on each iteration, the number (from 1 to 5) of modes in the distribution was determined. The goal is to determine the Bayes Factor (BF) for testing unimodality versus multimodality. If $a$ is the prior odds of more than one mode (determined by simulation in our code), and $b$ is the posterior odds of finding more than one mode, then
\begin{equation}
BF=b/a
\end{equation}
A $\log{BF} \leq{0} $ indicates that the distribution is unimodal. 
If $0 < \log(BF) < 1$, there is some weak evidence of multimodality \cite{Kass} (see Figure 1, A and D). If $1 < {\log(BF)} \leq{3}$, this is considered as positive evidence for multimodality. If $3 < \log(BF) \leq{5}$, this is considered as strong evidence for multimodality (see Figure 1, B and E). And, finally, a $\log(BF) >5$ is considered as very strong evidence for multimodality (see Figure 1, C and F). The reversible jump MCMC (rjMCMC) algorithm was used to fit the mixture model to each fixation via the “mixAK" R package \cite{Komarek}\footnote{R code for this computation is available at available at: \url{https://digital.library.txstate.edu/handle/10877/13485}}

\subsection{Finding stationary fixations}  
\label{sec:findstat}
Each fixation of data was assessed for stationarity. To illustrate the process, let us imagine that we are dealing with a fixation of 2314 samples. Each fixation was first divided into a first half and a second half by time.  So, the first half would begin at sample 1 and end at sample 1157. The second half would begin at sample 1158 and end at sample 2314.   These 2 subsections were evaluated for stationarity in the Zhivomirov and Nedelchev \cite{Stationarity} sense. That is, these segments were compared to determine if they had statistically significant different medians, variances, and autocorrelations. We also tested the first third vs the last 2 thirds (section 1 from samples 1 to 771, section 2 from 772 to 2314) and the first 2 thirds to the last third (section 1 from samples 1 to 1543, second section from 1544 to 2314). Also, we tested the first quarter to the last three quarters (section 1 from samples 1 to 578, section 2 from 579 to 2314) and from the first 3 quarters to the last quarter (section 1 from samples 1 to 1736, section 2 from 1737 to 2314).  So, each fixation was tested for stationarity with 5 splits. 

For all stationarity tests, mean stationarity was tested with a Mann-Whitney U-test and variance stationarity was tested with the Brown-Forsythe test.  

For autocorrelation stationarity, we obtained the lag 1, lag 2 and lag 3 temporal autocorrelation (acf1,  acf2, acf3) for each segment. We statistically compared the acf1, acf2 and acf3 values from each segment using an z-test \cite{cohen}.  If the p-values for any comparison were statistically significant ($p~<~0.05$, two-tailed) the fixations were considered non-stationary.

All of these data processing steps were applied to both the HA dataset and the TXstate dataset.  The key differences between the HA dataset and the TXstate dataset were that they were based on data from different subjects, and the HA dataset fixations consisted of approximately 50,000 samples collected at 500 Hz and the TXstate dataset fixations consisted of approximately 7,500 samples.  Also, they were collected from different laboratories using different eye-tracking models (EyeLink 1000+ vs EyeLink 1000). The extraordinary environmental requirements in the HA dataset were not applied to the TXstate dataset. 

\section{Results}
\label{sec:results}
\subsection{Checking Assumptions for Position Signals}
\subsubsection{Multimodality of Position Signals}
Table 2 provides a detailed analysis of our multimodality results for the position signals from both the HA dataset (top) and the TXstate dataset (bottom).  In the HA dataset, a substantial number, though not a majority of all fixations were multimodal.  The values ranged from 0 to 53\% multimodal with a median of 32\% of all fixations.  In the TXstate dataset, of 37 total fixations, 35 were multimodal.  
So, some fixations from the HA dataset are multimodal whereas for the TXstate dataset nearly all fixations were found to be multimodal.
It is interesting that even position data from artificial eyes can be multimodal.  Figure 2 illustrates an example of an artificial eye recording that is multimodal, non-stationary and temporally autocorrelated\footnote{For two additional examples, see  \url{https://digital.library.txstate.edu/handle/10877/15303}. Download ``Alternate\_Figures.zip''.}.

\begin{table*}[htbp]
\label{tab:tab2}
    \begin{adjustwidth}{-1.5in}{0in}
    \centering
    \begin{threeparttable}
    \caption{Multimodality and Stationarity of Position Signals and StoS Signals}
    \begin{tabular}{|c|c|c|c|c|c|c|}
        \multicolumn{7}{c}{\textbf{\large{HA Dataset - Human Data}}}\\ \hline
        Subject & Filter & N Segments & \% Multi (P) \tnote{1} & \% Multi (S) \tnote{2} & \% Stat (P) \tnote{3}& \% Stat (S) \tnote{4}\\ \hline
        KE & OFF & 17 & 35.29 & 0.00 & 0.00 & 52.94 \\ \hline
        SA & OFF & 33 & 6.06 & 0.00 & 0.00 & 24.24 \\ \hline
        KE & ON & 24 & 45.83 & 0.00 & 0.00 & 52.94 \\ \hline
        SA & ON & 38 & 52.63 & 0.00 & 0.00 & 24.24 \\ \hline
        \multicolumn{7}{c}{}\\
        \multicolumn{7}{c}{\textbf{\large{HA Dataset - Artificial Eye Data}}}\\ \hline
        Subject & Filter & N Segments & \% Multi (P) \tnote{1} & \% Multi (S) \tnote{2}& \% Stat (P) \tnote{3}& \% Stat (S)\tnote{4}\\ \hline
        KE & OFF & 17 & 0.00 & 0.00 & 0.00 & 33.33 \\ \hline
        SA & OFF & 33 & 36.36 & 0.00 & 0.00 & 23.68 \\ \hline
        KE & ON & 24 & 29.17 & 0.00 & 0.00 & 35.29 \\ \hline
        SA & ON & 38 & 15.79 & 0.00 & 0.00 & 42.42 \\ \hline
        \multicolumn{7}{c}{}\\
        \multicolumn{7}{c}{\textbf{\large{TXstate Dataset - Human Data}}}\\ \hline
        Subject & Filter & N Segments & \% Multi (P) \tnote{1}& \% Multi (S) \tnote{2}& \% Stat (P) \tnote{3}& \% Stat (S)\tnote{4}\\ \hline
        A & OFF & 2 & 100 & 0.00 & 0.00 & 0.00 \\ \hline
        A & ON & 5 & 100 & 0.00 & 0.00 & 20.00 \\ \hline
        B & OFF & 2 & 100 & 0.00 & 0.00 & 50.00 \\ \hline
        B & ON & 6 & 100 & 0.00 & 0.00 & 66.67 \\ \hline
        C & OFF & 2 & 100 & 0.00 & 0.00 & 0.00 \\ \hline
        C & ON & 7 & 100 & 0.00 & 0.00 & 28.57 \\ \hline
        D & OFF & 6 & 75 & 0.00 & 0.00 & 37.50 \\ \hline
        D & ON & 5 & 100 & 0.00 & 0.00 & 20.00 \\ \hline
    \end{tabular}
    \begin{tablenotes}
        \item[1] Percent Multimodal Position Signals
        \item[2] Percent Multimodal StoS Signals
        \item[3] Percent Stationary  Position Signals
        \item[4] Percent Stationary StoS Signals
    \end{tablenotes}
    \end{threeparttable}
    \end{adjustwidth}
\end{table*}

\begin{figure*}[htbp]] 
\label{fig:fig02}
\begin{adjustwidth}{-2in}{0in}\centering
\centering
    \includegraphics[width=1.0\linewidth]{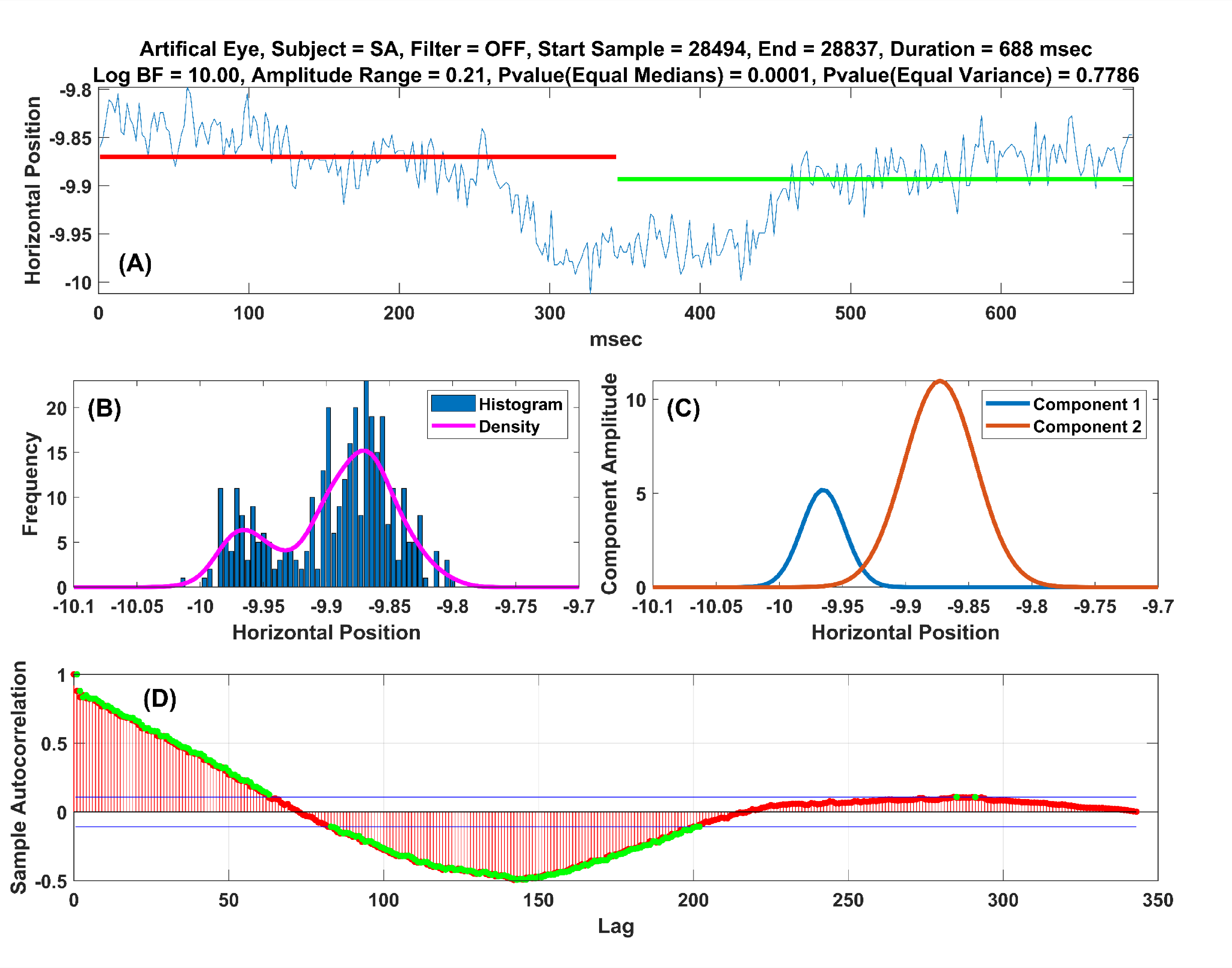}
\caption{Detailed analysis of a multimodal, non-stationary, and highly autocorrelated fixation from an artificial eye.  The data were calibrated using the same calibration as for the human subject SA in the filter off condition.  (\textbf{A}) The horizontal position signal for this fixation.  It is 688 msec in duration (344 samples).  Note the very small range of the data (~0.2 degrees).  The red line is the median for the first half of the data.  The green line is the median for the second half of the data.  The p-value for the test that the medians are equal is 0.0001 (the smallest p-value we allow). Therefore, we must reject the hypothesis that the medians are equal. Therefore, this segment is not stationary.  (\textbf{B})  A histogram of horizontal position and a histogram density plot.  The log(BF) was 10.0, which was the maximum log(BF) that we allowed.  Thus, there is very strong evidence that the distribution is multimodal (i.e., bimodal in this case). Our algorithm found 2 components.  (\textbf{C}) A plot of the two normal components that our program fit to this data.  (\textbf{D}) The autocorrelation function for the maximum allowable lags (1 to 343).  The green points were statistically significant ($p~<~0.05$).  Note that we still have statistically significant autocorrelation out to lags in the 280 range.}
\label{fig:fig2}
\end{adjustwidth}
\end{figure*}

\subsubsection{Stationarity of Position Signals}
\label{sec:statpos}
Table 2 provides a detailed analysis of our stationarity results for the position signals.  No position signals from any fixation from either dataset was stationary.

\subsubsection{Autocorrelation of Position Signals}
\label{sec:ACF-Position}
Every position signal from every fixation from either study had a statistically significant temporal autocorrelation ($p~<~0.01$).  The autocorrelation results for position signals are illustrated in Figure 3. Note that, in the HA study, there appeared to be little difference between the autocorrelation of human and artificial eye data.  For both the HA dataset and the TXstate dataset, unfiltered data had a lower ACF1 than that for filtered data, although this effect was much smaller for the TXstate dataset.

\begin{figure*}[htbp] 
\label{fig:fig3}
\centering
    \includegraphics[width=1.0\linewidth]{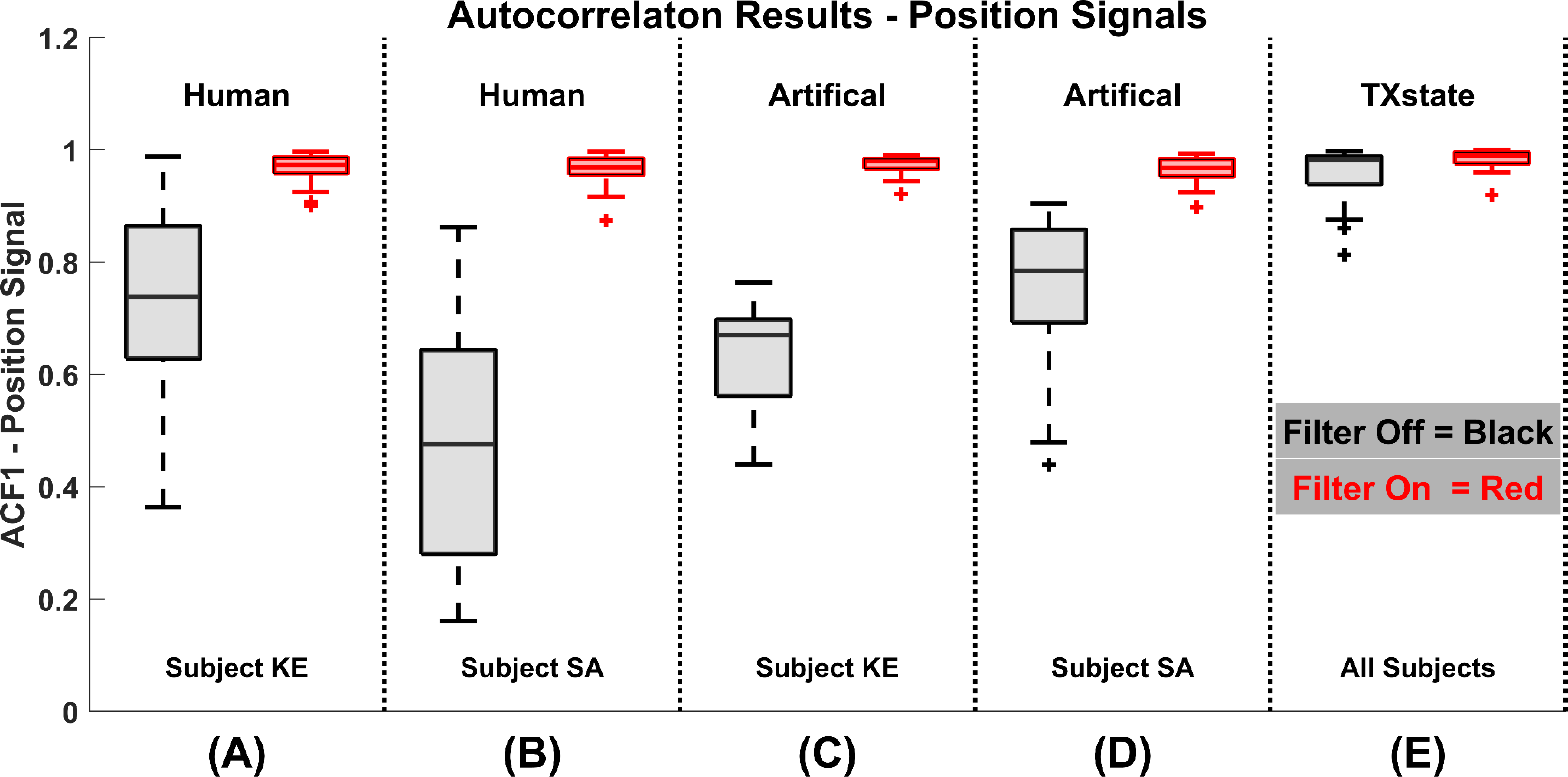}
\caption{Autocorrelation results (lag 1) for position signals. (\textbf{A}) HA dataset, Subject KE, Human data.  (\textbf{B}) HA dataset, Subject SA, Human data. (\textbf{C}) HA dataset, Subject KE, Artificial Eye data.  (\textbf{D}) HA dataset, Subject SA, Artificial Eye data. (\textbf{E}) TXstate dataset, all 4 subjects combined, Human data.}
\end{figure*}

Figure 4 (B and D) present autocorrelation results for position data out to 300 lags.  Note the presence of statistically significant autocorrelation out to 300 lags for unfiltered position data (B) as well as filtered position data (D).  In unfiltered position data, the ACF oscillates above and below 0.0 The filtered data stays positive up to about lag 140 and turns negative at around lag 220.

\begin{figure*}[htbp]] 
\label{fig:fig04}
\begin{adjustwidth}{-2in}{0in}\centering
\centering
    \includegraphics[width=1.0\linewidth]{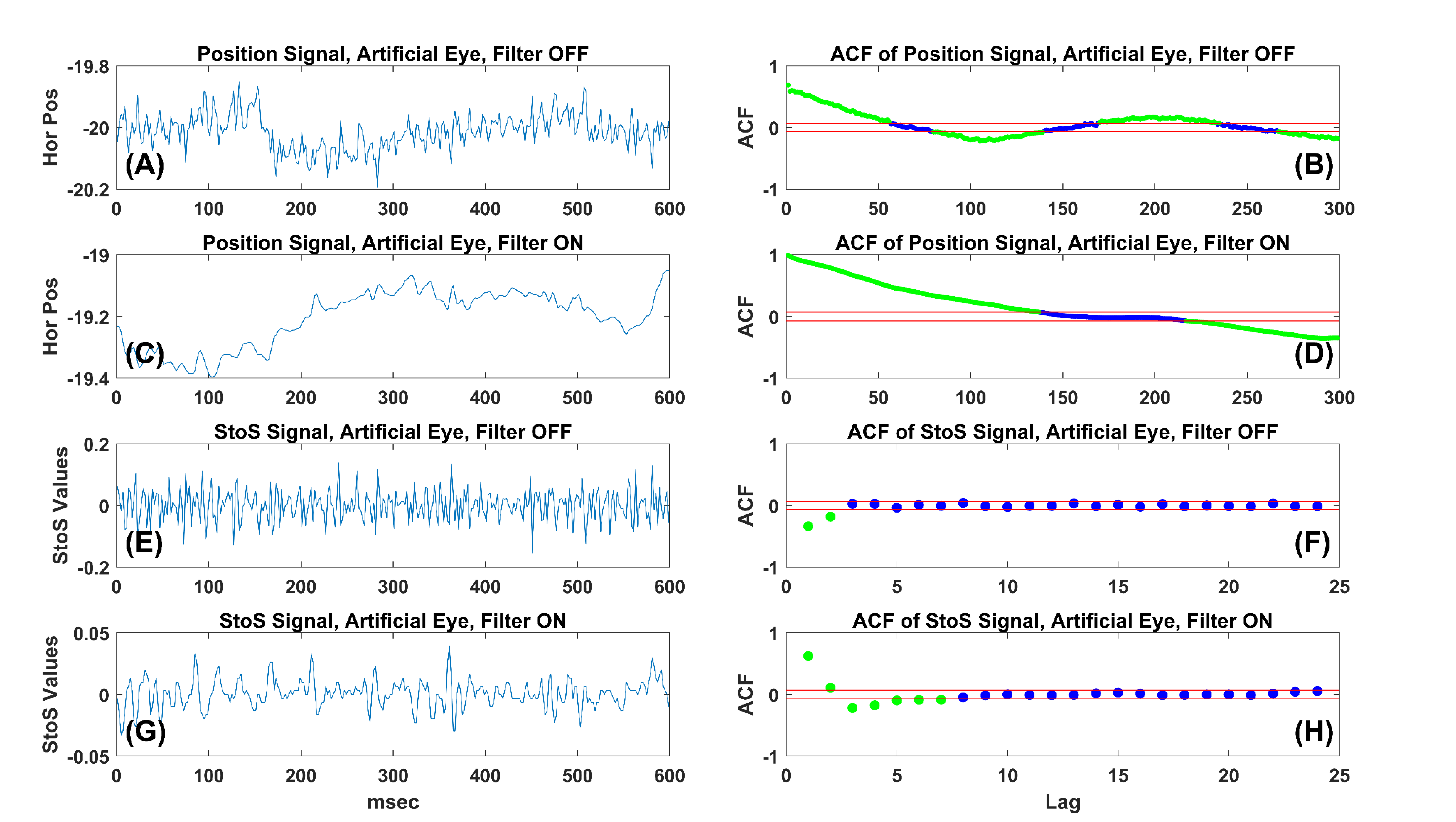}
\caption{Illustration of autocorrelation functions for artificial eyes. Lags of either 300 (position signals) or 25 (StoS signals) were employed.  On the left (A, C, E, G) are the first 600 sec of the fixation being tested.  On the right (B, D, F, H) are the multiple lag autocorrelation functions for each fixation plot on the left.  Green points have statistically significant autocorrelation values ($p~<~0.012$, two-tailed).}
\end{adjustwidth}
\end{figure*}

\subsection{Checking Assumptions for StoS Signals}
\subsubsection{Multimodality of StoS Signals}
Table 2 provides a detailed analysis of the multimodality results for StoS signals. All StoS signal distributions for both datasets were unimodal.

\subsubsection{Stationarity of StoS Signals}
Table 2 provides a detailed analysis of the stationarity of StoS signals.   Many of these StoS fixations were stationary. 

Figure 5 illustrates the range of percent of StoS fixations that were stationary for all subjects from both studies.  The histogram summarizes 16 cases from both datasets together (HA: 2 subjects * 2 filter conditions * 2 eye types = 8; TXstate: 4 subjects * 2 filter conditions = 8).   A median of 31\% of all StoS fixation signals were stationary.   It may be surprising that StoS signals from artificial eyes could be non-stationary.  Figure 6 presents four examples of StoS fixations from artificial eyes whose 50\% - 50\% splits were not variance stationary.

\begin{figure*}[htbp]] 
\label{fig:fig05}
\centering
    \includegraphics[width=1.0\linewidth]{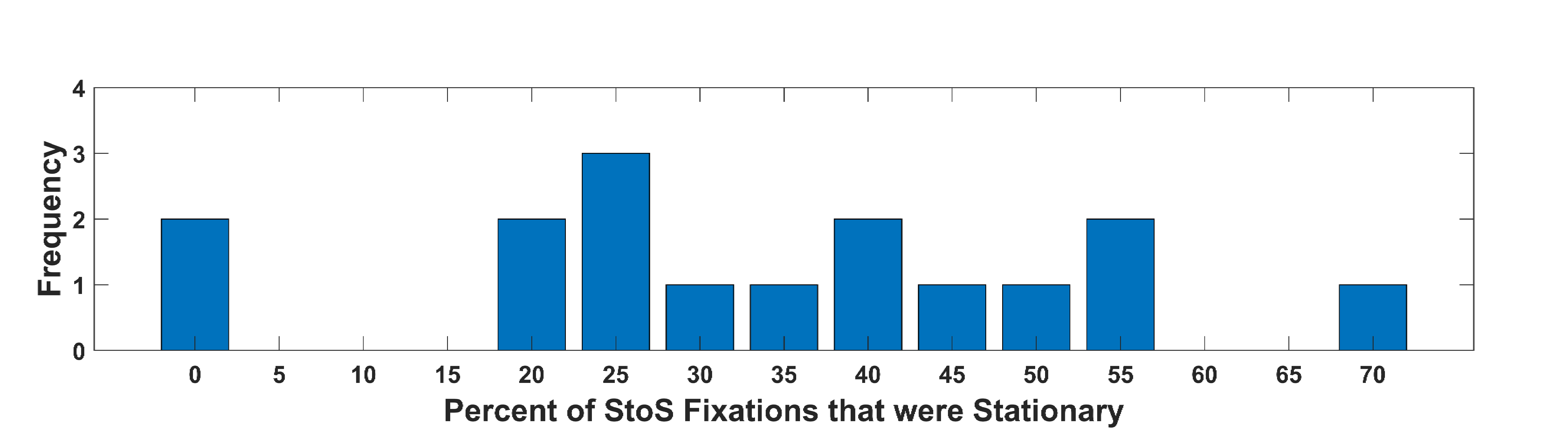}
\caption{Histogram of percent of StoS signals that were stationary.}
\end{figure*}


\begin{figure*}[htbp]] 
\label{fig:fig06}
\begin{adjustwidth}{-2in}{0in}\centering
    \includegraphics[width=1.0\linewidth]{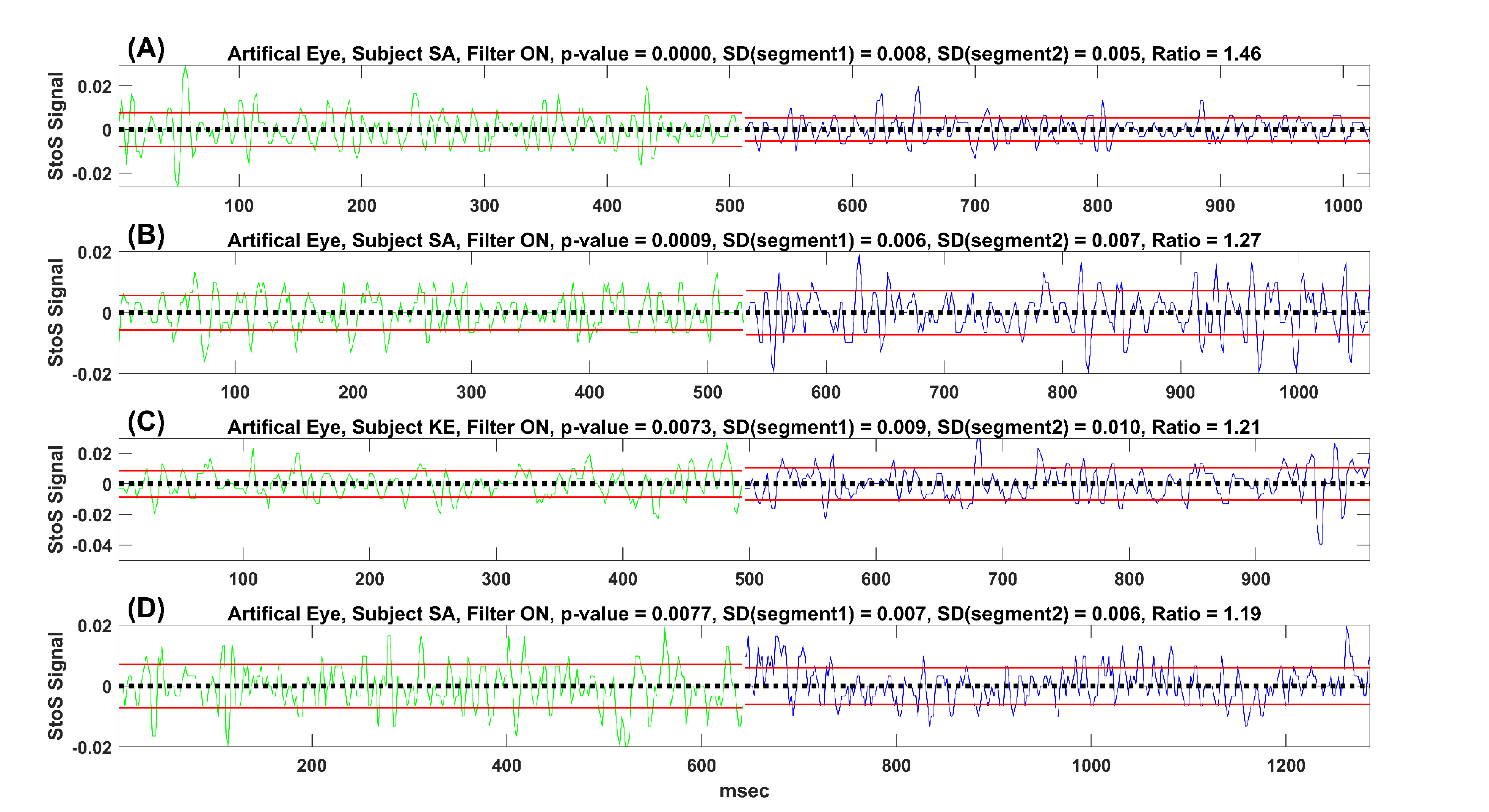}
\caption{Four examples of StoS fixations from artificial eyes that are not variance stationary.   All fixation tested for variance stationarity using a 50\% - 50\% segment length split.  (\textbf{A}) Segment of artificial eye data from subject SA in the filter ON condition.  The red lines are $\pm 1~SD$.  The p-value is from the Brown-Forsythe test of homogeneity of variance, out to 4 decimal places.   The SDs for the first and second half of the segment are provided.  Ratio is (highest SD)/ (lowest SD).  All other figures (B to D) follow same convention as (\textbf{A}).}
\end{adjustwidth}
\end{figure*}

Figure 4 (B and D) present autocorrelation results for position data out to 300 lags.  Note the presence of statistically significant autocorrelation out to 300 lags for unfiltered position data (B) as well as filtered position data (D).  In unfiltered position data, the ACF oscillates above and below 0.0 The filtered data stays positive up to about lag 140 and turns negative at around lag 220.

\subsection{Checking Assumptions for StoS Signals}
\subsubsection{Multimodality of StoS Signals}
Table 2 provides a detailed analysis of the multimodality results for StoS signals. All StoS signal distributions for both datasets were unimodal.

\subsubsection{Stationarity of StoS Signals}
Table 2 provides a detailed analysis of the stationarity of StoS signals.   Many of these StoS fixations were stationary. 

Figure 5 illustrates the range of percent of StoS fixations that were stationary for all subjects from both studies.  The histogram summarizes 16 cases from both datasets together (HA: 2 subjects * 2 filter conditions * 2 eye types = 8; TXstate: 4 subjects * 2 filter conditions = 8).   A median of 31\% of all StoS fixation signals were stationary.   It may be surprising that StoS signals from artificial eyes could be non-stationary.  Figure 6 presents four examples of StoS fixations from artificial eyes whose 50\% - 50\% splits were not variance stationary.

\subsubsection{Autocorrelation of StoS Signals}
In the HA dataset every StoS signal from every fixation had a statistically significant temporal autocorrelation ($p~<~0.01$). In the TXstate dataset, of 37 fixations tested, all but 7 had statistically significant temporal autocorrelation.  Of these 7, 3 were stationary. Since all StoS fixations were  unimodal, there were 3 fixations of 37 in the TXstate dataset that met all of our statistical assumptions.
The autocorrelation results for StoS signals are illustrated in Figure 7. Unfiltered StoS signals have an ACF1 that is below 0.0 whereas filtered StoS signals have an ACF greater than
0. The magnitude of the ACF1 of the StoS fixations were less than those for the position
fixations. Figure 4 (F) and (H) present autocorrelations up to lag 25 for artificial eyes
StoS fixations. For the StoS signal in Figure (E), filter off condition, the first two lags were statistically significantly negative.  For the StoS signal in Figure (G), filter on condition, the first 7 lags, both positive and negative, were statistically significant.

\begin{figure*}[htbp]] 
\label{fig:fig07}
\centering
    \includegraphics[width=1.0\linewidth]{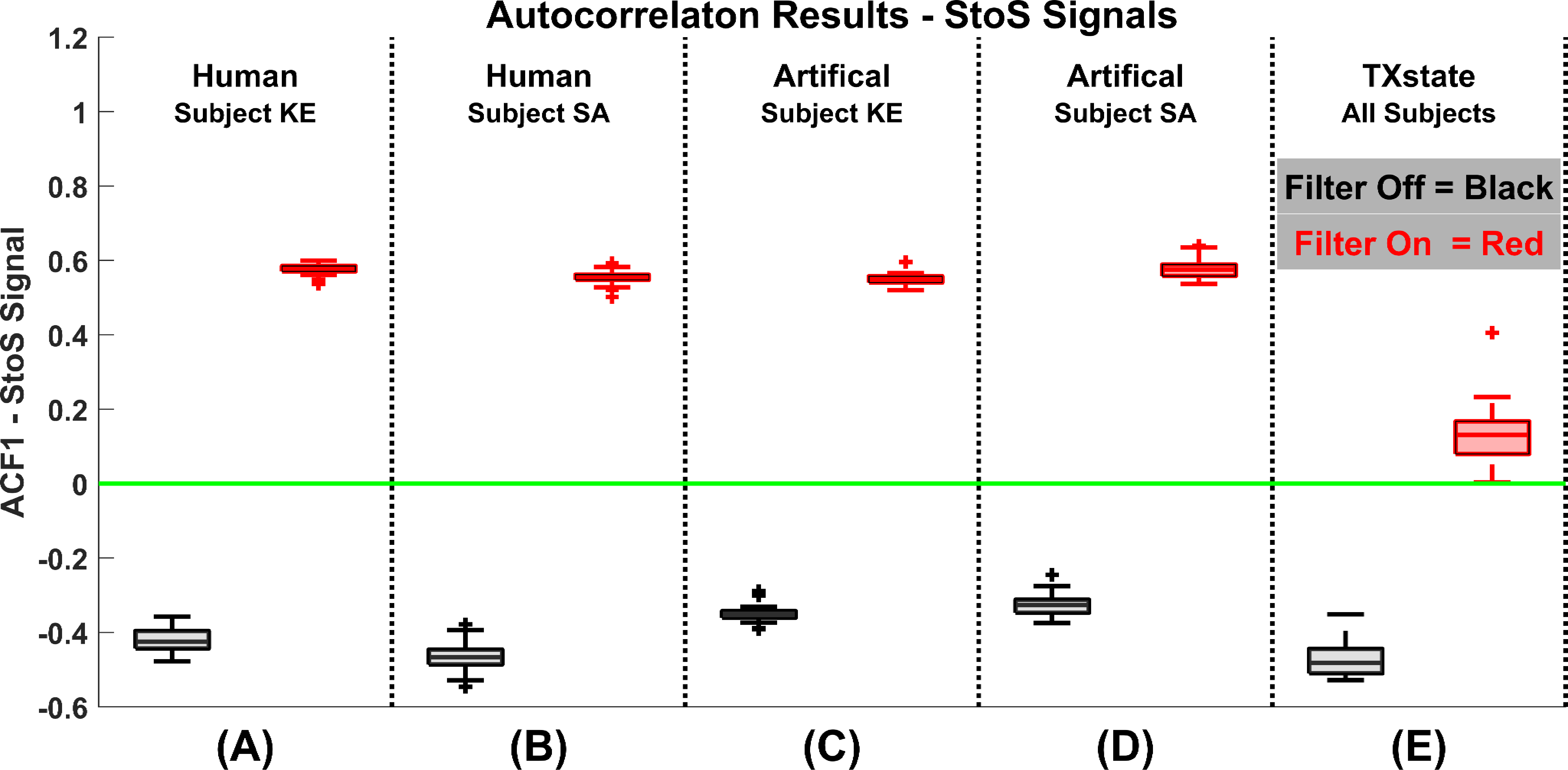}
\caption{Autocorrelation results (lag 1) for StoS signals. (A) HA dataset, Subject KE, Human data. (B)
HA dataset, Subject SA, Human data. (C) HA dataset, Subject KE, Artificial Eye data. (D) HA dataset,
Subject SA, Artificial Eye data. (E) TXstate dataset, all 4 subjects combined, Human data.
}
\end{figure*}

 \subsection{Checking Assumptions for Raw Signals}
Raw data (Figure 8), i.e., fixations in camera units prior to calibration, were autocorrelated.  All ACF1 values, for all the raw data fixations from every recording, for both horizontal and vertical position and StoS signals were statistically significant at the $p~<~0.001$ level. 

\begin{figure*}[htbp]] 
\label{fig:fig08}
\centering
    \includegraphics[width=1.0\linewidth]{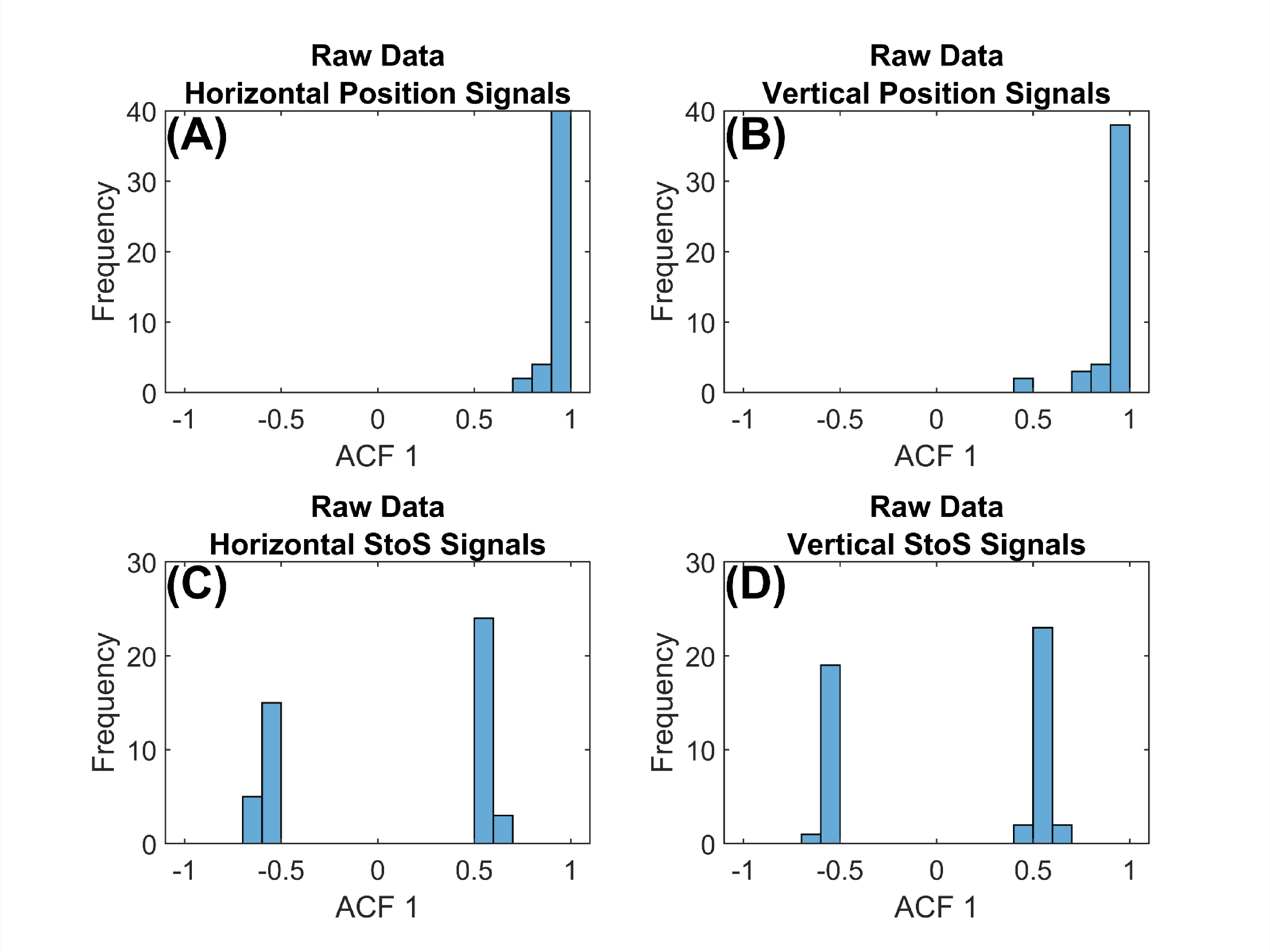}
\caption{Checking statistical assumptions in raw data from the TXstate dataset. (A) Histogram of ACF1
values for raw horizontal position signals. (B) Histogram of ACF1 values for raw vertical position signals.
  (C) Histogram of ACF1 values for raw horizontal StoS signals. There are two distributions, one below 0
from unfiltered data and one above 0 from filtered data. (D) Same as (C) for raw vertical StoS signals.
}
\end{figure*}

\section{Discussion}
\label{sec:Discuss}
The main findings of the present study are that the statistical assumptions underlying the use of the standard deviation (position signals) and RMS (StoS signals) as measures of precision are generally not met in time series of eye-movement fixation measured with the EyeLink 1000.  All of the position signals from both studies failed our tests of the three assumptions.  Of 261 StoS fixations tested from two datasets, only 3 fixations satisfied all three statistical assumptions.  These statements apply equally to recordings from human eyes and artificial eyes.

Position signals frequently had multimodal distributions, were never stationary and, in every case, exhibited statistically significant temporal autocorrelation.  StoS signals were always unimodal, were frequently non-stationary and, in almost every case, were statistically significantly temporally autocorrelated.  In the interest of keeping the manuscript within a reasonable length, all of our observations were based on horizontal gaze of the dominant eye.  We think it is reasonable to assume that other signals (vertical, non-dominant eye) would show a similar pattern, but it is an empirical question.

Temporal autocorrelation was similar for human and artificial eyes.  This points to a minor role of human characteristics in the autocorrelation result.  It would seem that the simple act of recording “eye movement data” using the video-oculography method, and with no filtering, still results in moderate to high levels of temporal autocorrelation.  In this light, it is interesting to note that even “raw” data signals were also autocorrelated.

Some position signals were multimodal, and some were not.  Although multimodality was distinctly less of an issue for the super-fixators in the HA dataset, generally, approximately 30\% of all fixations were multimodal.  It was also interesting to note that a number of fixations from artificial eyes exhibited multimodality. So, multimodality was not strictly a function of human characteristics.

No position signal fixations were found to be stationary, although many of our StoS signal fixations were stationary.  It would be reasonable to characterize our method of determining stationarity as strict, especially considering the number of splits that we employed.  We note that our tests of stationarity for fixations appear to have very large statistical power.  This would markedly decrease the probability of finding stationarity.

Blignaut and Beelders \cite{Blignaut} propose the use of the Bivariate Contour Ellipse Area (BCEA) for measurement of the precision of eye-trackers.  If one plots the horizontal and vertical eye position sample data in a 2D plot, the data should form some sort of cluster.  The idea of the BCEA is to fit an ellipse to that cluster and determine the elliptical area that would encompass 63.2\% of all sample pairs.  This method assumes unimodality.  Although multimodality was not an overwhelming issue with the two “fixators” from the HA study, there were still approximately 30\% of fixations that were multimodal.    In the TXstate study, almost all fixations were multimodal.  We don’t think it is appropriate to use the BCEA in the presence of multimodality.  Castet and Crossland \cite{Castet} provide an approach to quantification in the presence of spatial multimodality.

The SD calculates the typical deviation around the mean (or central value) of a set of observations.  If the distribution of the observations is multimodal then the mean of the distribution is not likely to serve as a useful summary of the center of the distribution and the usual standard deviation can provide a misleading picture of the variation in the data.  The same issue arises if the series of observations is not stationary because the level or variation changes over the length of the series.  Finally, computing the standard deviation for data that are not independent, e.g., data that have nonzero autocorrelations, is not appropriate because the usual sample standard deviation is estimating a complex function that depends on the variance and correlation structure of the sequence.  
Our evidence indicates that the application of the SD to position signals and the application of the RMS to StoS signals are generally not formally appropriate for eye fixation data.  In our next paper, we intend to propose statistics to replace these metrics with metrics that serve the same purpose but do not violate underlying assumptions.

\section*{Acknowledgments}
The study was funded by 3 grants to Komogortsev: (1)~National Science Foundation, CNS-1250718 and CNS-1714623, www.NSF.gov; (2)~National Institute of Standards and Technology, 60NANB15D325, www.NIST.gov; (3)~National Institute of Standards and Technology, 60NANB16D293. Stern’s work was partially funded by the Center for Statistics and Applications in Forensic Evidence (CSAFE) through Cooperative Agreement 70NANB20H019 between NIST and Iowa State University, which includes activities carried out at Carnegie Mellon University, Duke University, University of California Irvine, University of Virginia, West Virginia University, University of Pennsylvania, Swarthmore College and University of Nebraska, Lincoln.

\section*{Conflict of interest}
The authors declare that they have no conflict of interest.

\section*{Open Practices Statement} As stated in the manuscript, the signals for all fixations analyzed are at  \url{https://digital.library.txstate.edu/handle/10877/15303}.  Examples of manually removed non-fixation events is also available at that site.  Also, alternate Figure 2 examples are available there.  The R-code for assessing multimodality is available at \url{https://digital.library.txstate.edu/handle/10877/13485}

\bibliography{library}

\bibliographystyle{abbrv}

\end{document}